\documentclass[12pt]{article}
\usepackage{graphicx}
\usepackage{tikz}
\usetikzlibrary{trees}
\usetikzlibrary{decorations.pathmorphing}
\usetikzlibrary{decorations.markings}
\usetikzlibrary{arrows}

\def\oxford{CERN, Geneva, Switzerland}

\def\pbnr{}
\def\Title#1{\begin{center} {\Large #1 } \end{center}}
\def\Author#1{\begin{center}{ \sc #1} \end{center}}
\def\Address#1{\begin{center}{ \it #1} \end{center}}

\newcommand\pubnumber{\pbnr}
\newcommand\pubdate{\today}
\def\Title#1{\begin{center} {\Large #1 } \end{center}}
\def\Author#1{\begin{center}{ \sc #1} \end{center}}

\newcommand{\OnBehalf}[1]{\sbox0{#1}\ifdim\wd0=0pt
        {}
	\else
	{\\on behalf of #1}
	\fi}
\newcommand{\SupportedBy}[1]{\sbox0{#1}\ifdim\wd0=0pt
        {}
	\else
	{\footnote{#1}}
	\fi}
\def\Address#1{\begin{center}{ \it #1} \end{center}}

\newcommand\pubblock{\includegraphics[width=5cm]{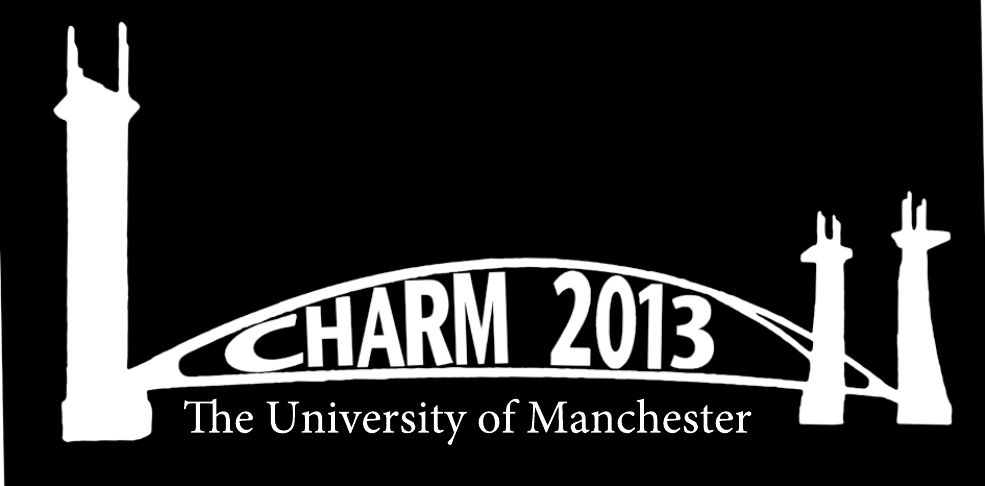}\hfill{\begin{tabular}{l} \pubnumber\\
         \pubdate  \end{tabular}}}
\newenvironment{Abstract}{\begin{quotation}  }{\end{quotation}}
\newenvironment{Presented}{\begin{quotation} \begin{center} 
             PRESENTED AT\end{center}\bigskip 
      \begin{center}\begin{large}}{\end{large}\end{center} \end{quotation}}

\def\beq{\begin{equation}}
\def\eeq#1{\label{#1}\end{equation}}
\def\eeqn{\end{equation}}

\def\beqa{\begin{eqnarray}}
\def\eeqa#1{\label{#1}\end{eqnarray}}
\def\eeqan{\end{eqnarray}}

\let\bar=\overbar

\def\Dslash{\not{\hbox{\kern-4pt $D$}}}
\def\dslash{\not{\hbox{\kern-2pt $\del$}}}

\def\msb{{\bar{\ssstyle M \kern -1pt S}}}

\begin{document}
\begin{titlepage}
\pubblock

\vfill
\Title{Measurement of production asymmetries
 }
\vfill
\Author{ Hamish Gordon, for the LHCb Collaboration}
\Address{\oxford}
\vfill
\begin{Abstract}
The knowledge of charm production asymmetries is an important prerequisite
for many of the possible searches for CP violation in
charm. Measurements of these asymmetries at hadron colliders can also help to improve our
understanding of QCD. These proceedings review existing measurements
and discuss some of the experimental challenges of determining
charge asymmetries at the per-mille level.

\end{Abstract}
\vfill
\begin{Presented}
The 6th International Workshop on Charm Physics\\
(Charm 2013)\\
29th August-3rd September, Manchester, UK
\end{Presented}
\vfill
\end{titlepage}
\def\thefootnote{\fnsymbol{footnote}}
\setcounter{footnote}{0}

\section{Introduction}

The recent hints of CP violation (CPV) in singly Cabibbo suppressed $D^{0}$ decays to two-body
final states from LHCb~\cite{charmcpv} and
CDF~\cite{Collaboration:2012qw} have heightened interest from
theoreticians in charm physics. Despite the lack of confirmation of
these hints by further studies~\cite{dACPSL}, searches for direct CPV in
charm remain well motivated. Measurements of charm production
asymmetries have the potential to increase the number of possible
techniques for CPV searches in charm, and also to make existing
searches more precise. 

For example, the most powerful search technique is currently the measurement of the
$\Delta A_{CP}$ observable, which is the difference in
CP-violating asymmetries between $D^0 \to K^-K^+$ and $D^0 \to
\pi^-\pi^+$. This quantity is equal to the difference between the
measured raw asymmetries in these decay modes, where the raw asymmetry
is defined for observed numbers of decays $N$ as
\begin{equation}
    A_{raw} = \frac{N(D^0)-N(\overline{D^0})}{N(D^0)+N(\overline{D^0})}.
\end{equation}

The largest useable samples of these decays are those that originate
from $D^{*+}$ decays to $D^0$ and a charged pion, which tags the
flavour of the $D^0$. Unfortunately, the values of $\Delta A_{CP}$
expected in the Standard Model are difficult to calculate, partly due to the lack of a good understanding of the
strong interaction effects and partly because the charge asymmetries in
the individual decay modes are not known. Knowledge of the production
asymmetry in $D^{*+}$ decays would enable measurements of the charge
asymmetries in $D^0 \to K^-K^+$ and $D^0 \to
\pi^-\pi^+$ separately, solving the second of these
problems. Furthermore, a precise production asymmetry measurement
could in principle lead to a more precise measurement of CP violation in $D^0 \to
K^-K^+$  than that in $\Delta A_{CP}$, where the statistical
uncertainty is limited by the $D^0 \to
\pi^-\pi^+$ decay channel. Measurements of production asymmetries in
the $D^+$, $D_s^+$ and $\Lambda_{c}^+$ sectors
are also worthy endeavours which will pave the way for more precise
searches for CP violation in their Cabibbo-suppressed decay modes.

 Measurements of charm production asymmetries are also interesting in
 their own right. The huge samples of charm decays from proton-proton collisions available at the LHC
 experiments can be used to improve our knowledge of the structure of the proton. It
 is conceivable that the charm samples at the B-factories could also
 be used to make precise tests of QCD symmetries via an investigation of
 the foward-backward asymmetry in charm meson production.

The forward-backward asymmetry in $D^{\pm}$ production has
been measured at the Belle experiment, and the $D^+$ and $D_s^+$
production asymmetries in $pp$ collisions have been measured at
LHCb. These measurements, discussed in the next sections, all use large Cabibbo-favoured charm
samples, in which no CP violation is expected. To make full use of
the high statistical precision possible with these samples, careful
studies of the systematic effects intrinsic to charge asymmetry
measurements in particle physics detectors are required, and these are also
discussed here.

\section{Production asymmetry measurements at $e^+e^-$ colliders}

 In the search for CPV in $D^+ \to K^0_S\pi^+$ at the Belle experiment~\cite{BelleKsPi}, the CP, detector and
  production asymmetries are intertwined. The raw measured charge asymmetry is
\begin{eqnarray}
\nonumber
  A^{K^0_S\pi^+}_{\rm rec}&=&A^{K^0_S\pi^+}_{CP}~+~A^{D^+}_{FB}(\cos\theta^{\rm CMS}_{D^+})\\
  &+&A^{\pi^+}(p^{\rm lab}_{T\pi^+},\cos\theta^{\rm
  lab}_{\pi^+})~+~A_{\mathcal{D}}(p^{\rm lab}_{K^0_S})
  \label{EQ:ARECONII}
\end{eqnarray}
where $A^{\pi^+}(p^{\rm lab}_{T\pi^+},\cos\theta^{\rm
  lab}_{\pi^+})$ and $A_{\mathcal{D}}(p^{\rm lab}_{K^0_S})$ are the
detection asymmetries of charged pions and neutral kaons respectively. The quantities $p^{\rm lab}$ and
$p^{\rm lab}_{\rm T}$ refer to momentum and transverse momentum in the
laboratory frame. The angle $\theta$ is the angle of the pion with respect
to the axis of the beam, in either the laboratory (lab) frame or the
centre of mass (CMS) frame. $A^{D^+}_{FB}(\cos\theta^{\rm CMS}_{D^+})$
is the forward-backward production
  asymmetry. $A^{\pi^+}(p^{\rm lab}_{T\pi^+},\cos\theta^{\rm
  lab}_{\pi^+})$ is measured as the difference in raw charge
asymmetries between $D^+ \to K^-\pi^+\pi^+$ and $D^0\to K^-\pi^+\pi^0$
under the assumption that $D^0$ and $D^+$ have the same forward-backward asymmetry.
$A_{\mathcal{D}}(p^{\rm lab}_{K^0_S})$ is calculated as discussed in
Sect.~\ref{sec:sys}, and subsequently subtracted.

The asymmetries due to production and CP violation are then determined. In terms of the charge asymmetry after correction for detector
  effects, $A^{K^0_S\pi^+_{\rm corr}}_{\rm rec}$, they are
\begin{eqnarray}
  \nonumber
 A^{K^0_S\pi^+}_{CP}&=&[A^{K^0_S\pi^+_{\rm corr}}_{\rm rec}(+\cos\theta^{\rm CMS}_{D^+})\\    
    &+&~A^{K^0_S\pi^+_{\rm corr}}_{\rm rec}(-\cos\theta^{\rm CMS}_{D^+})]/2,\\
  \nonumber
   A^{D^+}_{FB}&=&[A^{K^0_S\pi^+_{\rm corr}}_{\rm rec}(+\cos\theta^{\rm CMS}_{D^+}) \\
    &-&~A^{K^0_S\pi^+_{\rm corr}}_{\rm rec}(-\cos\theta^{\rm CMS}_{D^+})]/2
\end{eqnarray}
respectively. The CP asymmetry is consistent with CPV in the neutral kaon
  system as expected: $D^+ \to K^0_S\pi^+$ is a predominantly
  Cabibbo-favoured decay with no loop contribution at first
  order. 
\begin{figure}[htb]
\begin{center}
\includegraphics[width=8cm]{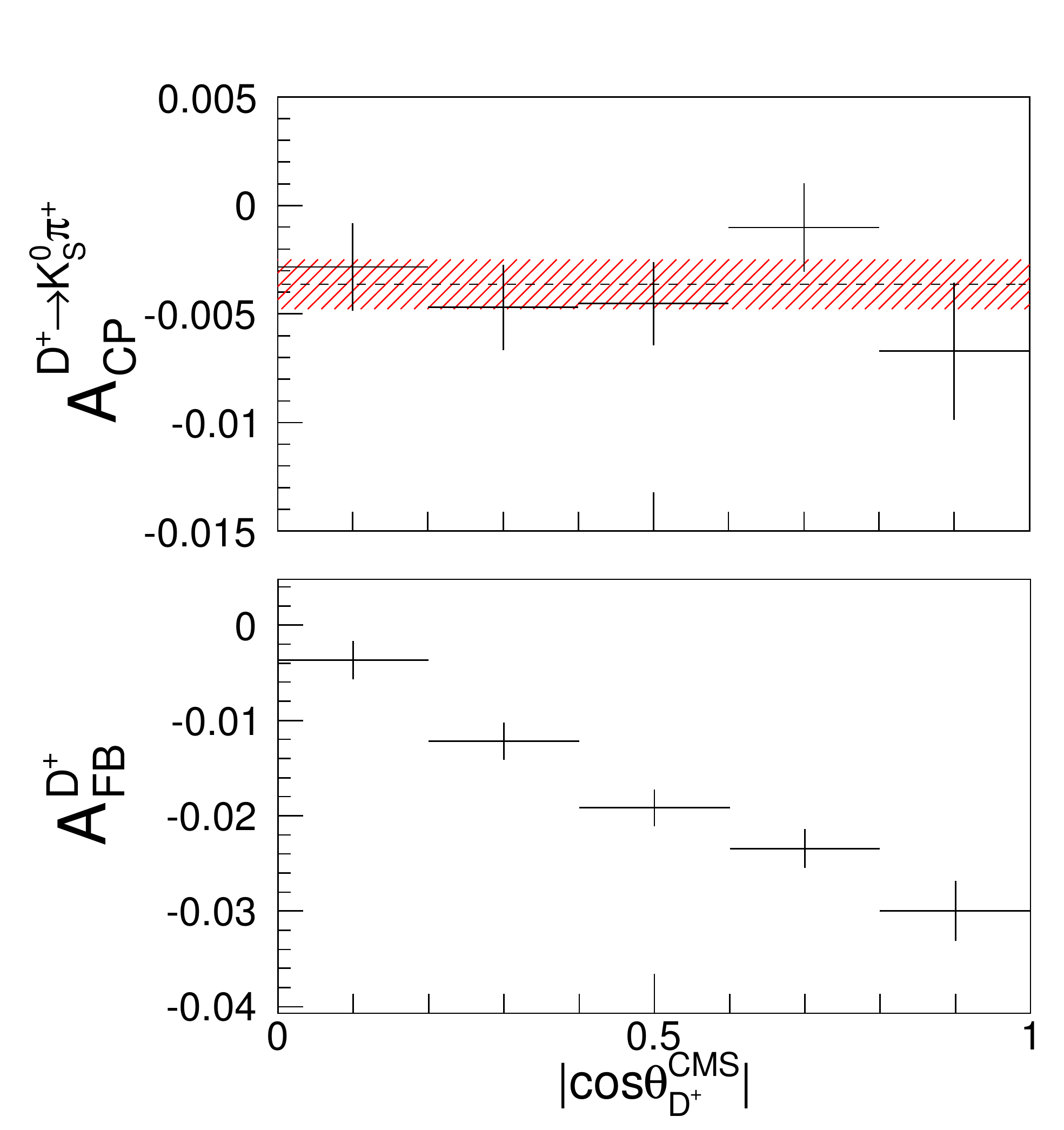}
\end{center}
\caption{ CP (top) and production (bottom) asymmetries measured in the $D^+ \to
  K^0_S\pi^+$ channel with a data sample corresponding to an
  integrated luminosity of 977 fb$^{−1}$ at the Belle experiment~\cite{BelleKsPi}. \label{fig:lb:massdp}}

\end{figure}

The assumption that the forward-backward asymmetries
 in charm meson production at $e^+e^-$ colliders does not depend on the
 flavour of the other quark in the meson could be tested if the pion
 efficiency asymmetry could be determined using another method,
 for example the technique outlined in Sect.~\ref{sec:sys} that has been
 employed at LHCb.

\section{Production asymmetries at hadron colliders}

When two protons collide, the baryon number conservation law implies
that two more baryons than antibaryons will form in the final
state. These will sometimes contain charm quarks, and thus one expects
an excess of charmed baryons over charmed antibaryons, with the effect
being more pronounced at high rapidity where the valence quarks tend
to end up. The anticharm quark formed with the charm quark must form
part of a meson, resulting in an excess of $\bar{D^0}$ over $D^0$ and
of $D^-$ over $D^+$. 
 It is helpful to define the Feynman momentum $x_F$ as the fraction
of the longitudinal momentum $p$ carried by the relevant parton. This $x_{F}$
is related approximately to rapidity $\eta$, transverse mass $m_{T}$ and
centre-of-mass energy $\sqrt{s}$ by
\begin{equation}
x_F \sim 2m_T e^{\eta}/\sqrt{s}
\end{equation}
  for $\eta > 1$. Since the valence quarks are found at high rapidity,
  the production asymmetry is likely to increase with $x_F$. 

In perturbative quantum chromodynamics (pQCD), charm quarks are produced by processes such as $q + \bar{q}
  \to c + \bar{c}$ and $g+g \to c + \bar{c}$,
with the second  dominating at high energy. Neither of these yield an overall
excess of one quark type over the other, however. Such net production
asymmetries cannot be explained with pQCD,  nor with the string fragmentation model
contained in the PYTHIA framework typically used to simulate $pp$
interactions at collider experiments. More creative explanations must
therefore be devised. Some models, for example the `meson cloud
model' \cite{mesoncloud}, assume that the incoming proton fluctuates into a virtual
charm meson - charm baryon pair which can sometimes escape and become real.
 Alternatively it has been proposed that $\bar{c}c$ pairs exist in the sea
 and have some probability to `recombine' with valence quarks and hadronise~\cite{recomb}.
 These two models lead to different forecasts for the energy
  dependence of the production asymmetry, and Ref.~\cite{mesoncloud}
  contains concrete predictions which are compared to LHCb
  measurements of the $D^{\pm}$ asymmetry.

\begin{figure}
\begin{center}
\includegraphics[width=7.2cm]{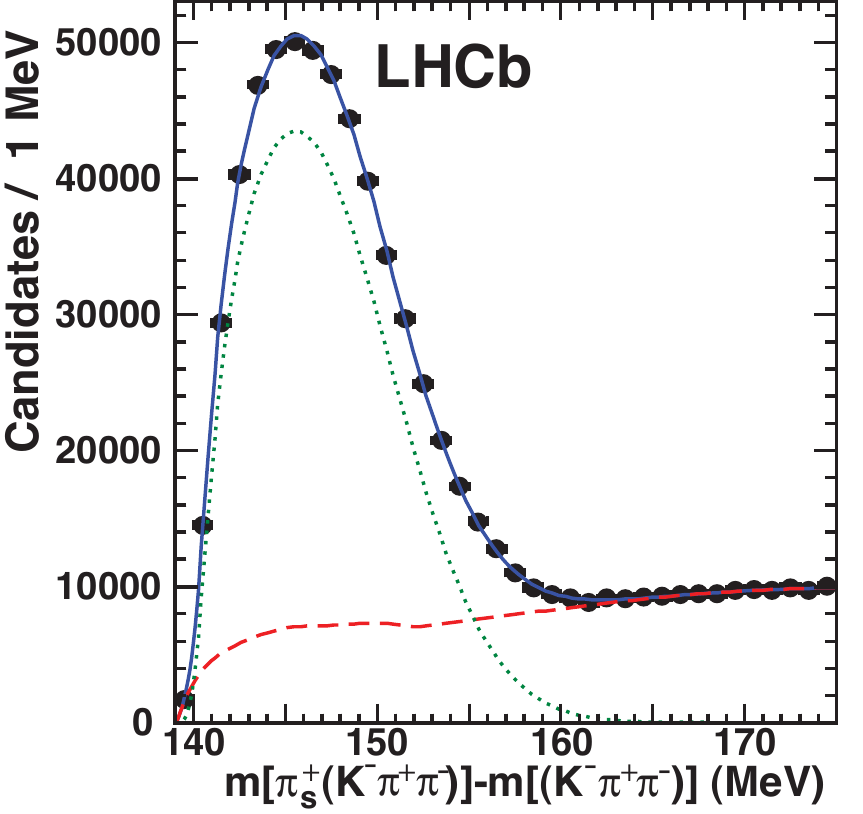}
\includegraphics[width=6.8cm]{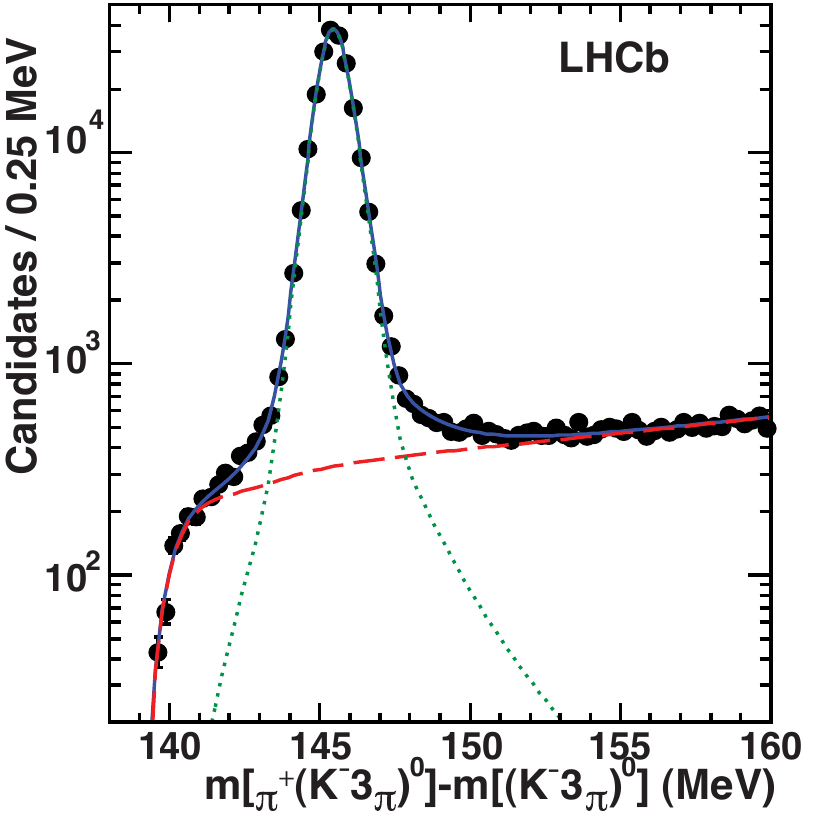}
\end{center}
\caption{Mass distributions for the tagging $D^*(2010)^+$ particle in
    the partially (left) and fully (right) reconstructed $D^0$ decays
    to $K^-\pi^+\pi^-\pi^+$.\label{fig:dstarmass}}
\end{figure}

The LHCb collaboration has measured both the $D^{\pm}$ and the $D_s^{\pm}$
production asymmetries~\cite{dsprod, dplusprod}. The $D_s^{\pm}$ asymmetry was determined using
$D_s^+ \to \phi\pi^+$ decays. In this case, the raw measured asymmetry
and the production asymmetry must be corrected by the detection asymmetry of the
charged pion. This was determined using tagged $D^0$ decays to
$K^-\pi^+\pi^-\pi^+$. Due to the large number of kinematic constraints
provided by the four-body final state, it is possible to reconstruct
$D^*(2010)^+$-tagged decays of this type with one pion missing. The pion tracking efficiency is then the yield of fully reconstructed
  decays divided by the yield of decays partially reconstructed with
  a missing pion.  The
  mass distributions for these two cases in the data recorded at LHCb in 2011 are shown in
  Fig.~\ref{fig:dstarmass}.

\begin{figure}
\begin{center}
\includegraphics[width=14cm]{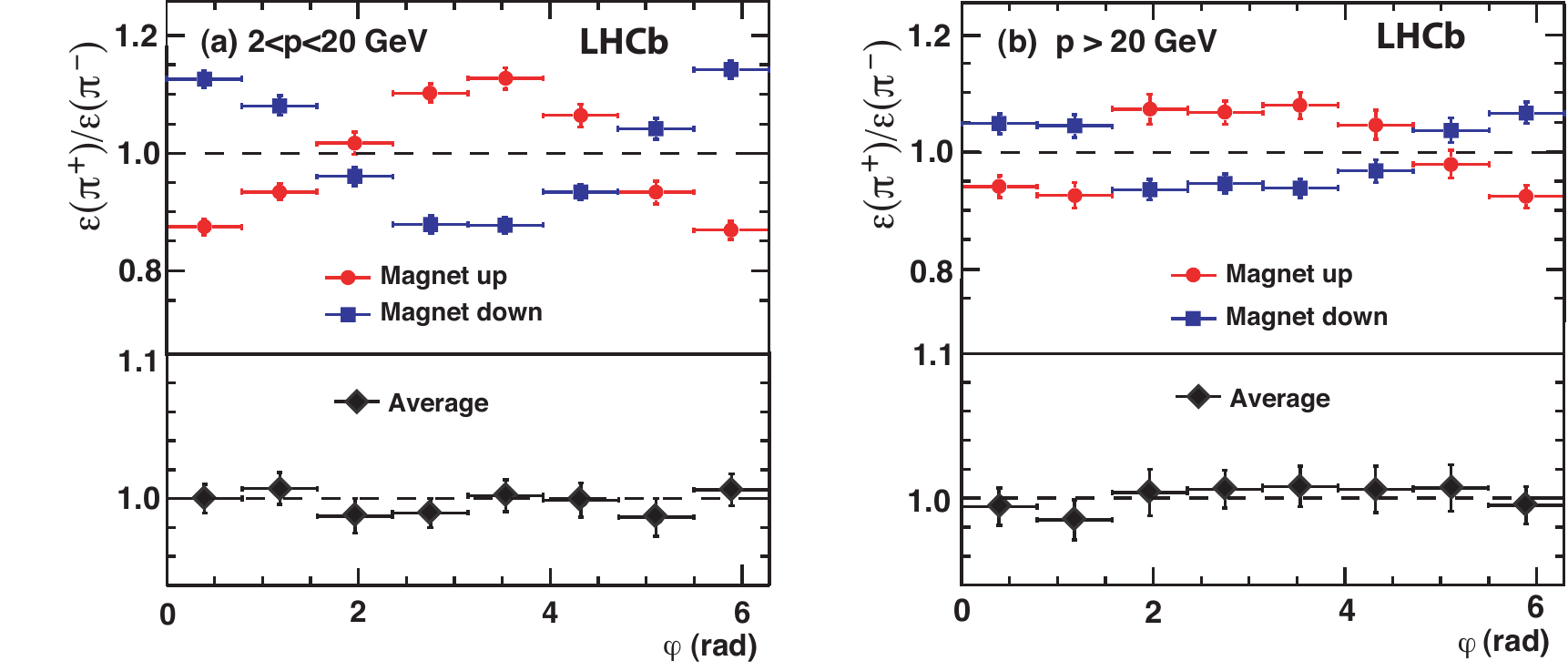}
\end{center}
\caption{The variation of the asymmetry in the pion tracking asymmetry
  with the azimuthal angle $\phi$ made by the pion with a horizontal
  plane defined across the centre of the detector, which is the bending plane of the
  magnet. The causes and ramifications of the variation in the data
  split according to the polarity of the magnet are
  discussed in Sec.~\ref{sec:sys}.\label{fig:pioneffic}}
\end{figure}

 The tracking efficiencies are determined for $D^{*+}$ and
  $D^{*-}$ separately, and thus the charge asymmetry in the pion
  detection efficiency is obtained. When averaged over the LHCb
  acceptance, the asymmetry is small, of order 0.1\%. However, for a
  given polarity of the LHCb magnet, the
  asymmetry varies quite strongly according to where in the detector the pions end
  up, as shown in Fig.~\ref{fig:pioneffic}. This is discussed further
  in Sect.~\ref{sec:sys}.

 In a similar analysis, the $D^+$ asymmetry was measured using $D^+ \to
K^0_S\pi^+$ decays. Here
\begin{equation}
   A_{prod} = A_{raw} - A_{\pi^+} -A_{K^0_S}
\end{equation} 
where the $K^0_S$ asymmetry $A_{K^0_S}$ is due to CP
violation and
material interactions in the neutral kaon system. There is assumed to be no CPV in
  the $D^+$ decay.

\begin{figure}
\begin{center}
\includegraphics[width=6.2cm]{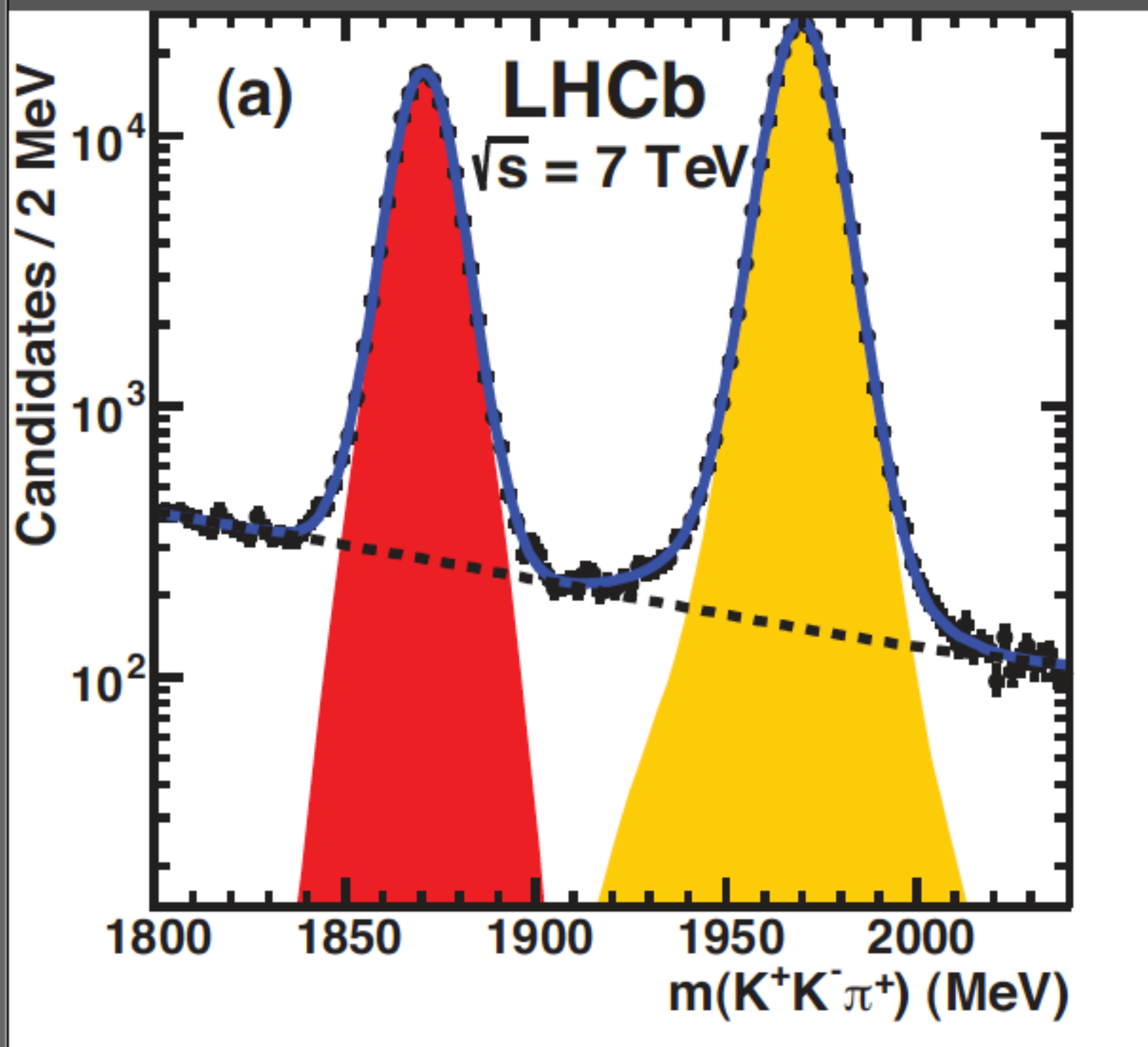}
\includegraphics[width=7.8cm]{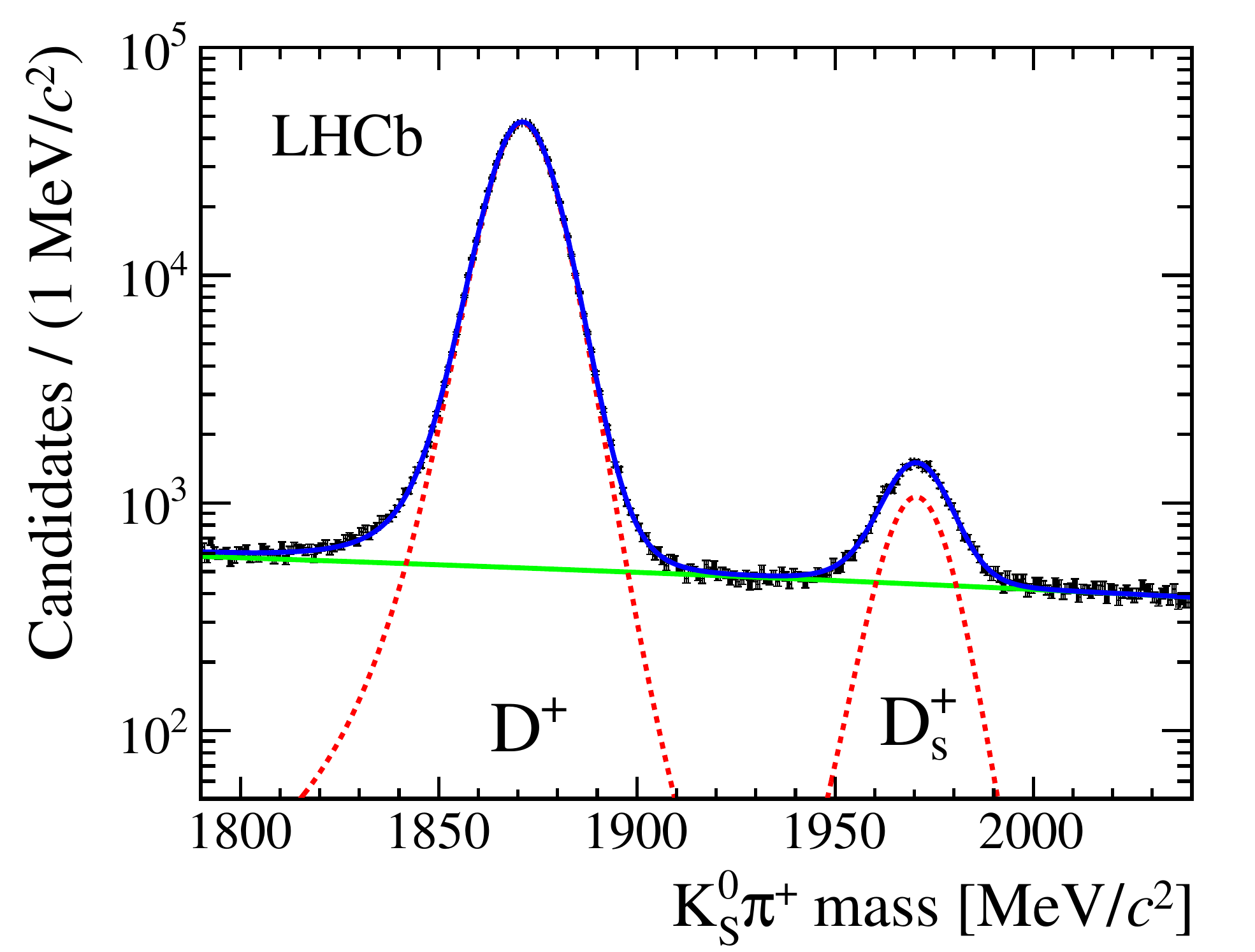}
\end{center}
\caption{Invariant mass distributions of the two final state particle combinations used to
  measure the $D^+_s$ (left) and $D^+$ (right) production
  asymmetries. In both cases, the $D^+_s$ and $D^+$ mass peaks are visible. \label{fig:masspeaks}}
\end{figure}

 To determine the production asymmetries, the yields of $D_{(s)}^+$ and $D_{(s)}^-$ decays, and the average pion efficiency asymmetries, are determined in $p_{\rm T}$ and $\eta$
    bins. The overall yields are shown in Fig.~\ref{fig:masspeaks}. The raw asymmetries are thus corrected for the pion
    asymmetry on a per-bin
    basis. Measured raw asymmetries in bins of $p_{\rm T}$ and $\eta$ are weighted
  by the reconstruction efficiency in these bins to determine an
  average asymmetry, and finally the charge asymmetry due to the
  neutral kaon is subtracted in the case of the $D^+$ measurement. This last quantity is very small because
  only neutral kaons with very short lifetimes are selected for use in
  the analysis, and thus its variation with $p_{\rm T}$ and $\eta$ is
  negligible. The results are asymmetries for $D_{(s)}^+$ decays
  produced in $pp$ collisions in the LHCb acceptance.

\begin{figure}
\begin{center}
\includegraphics[width=7cm]{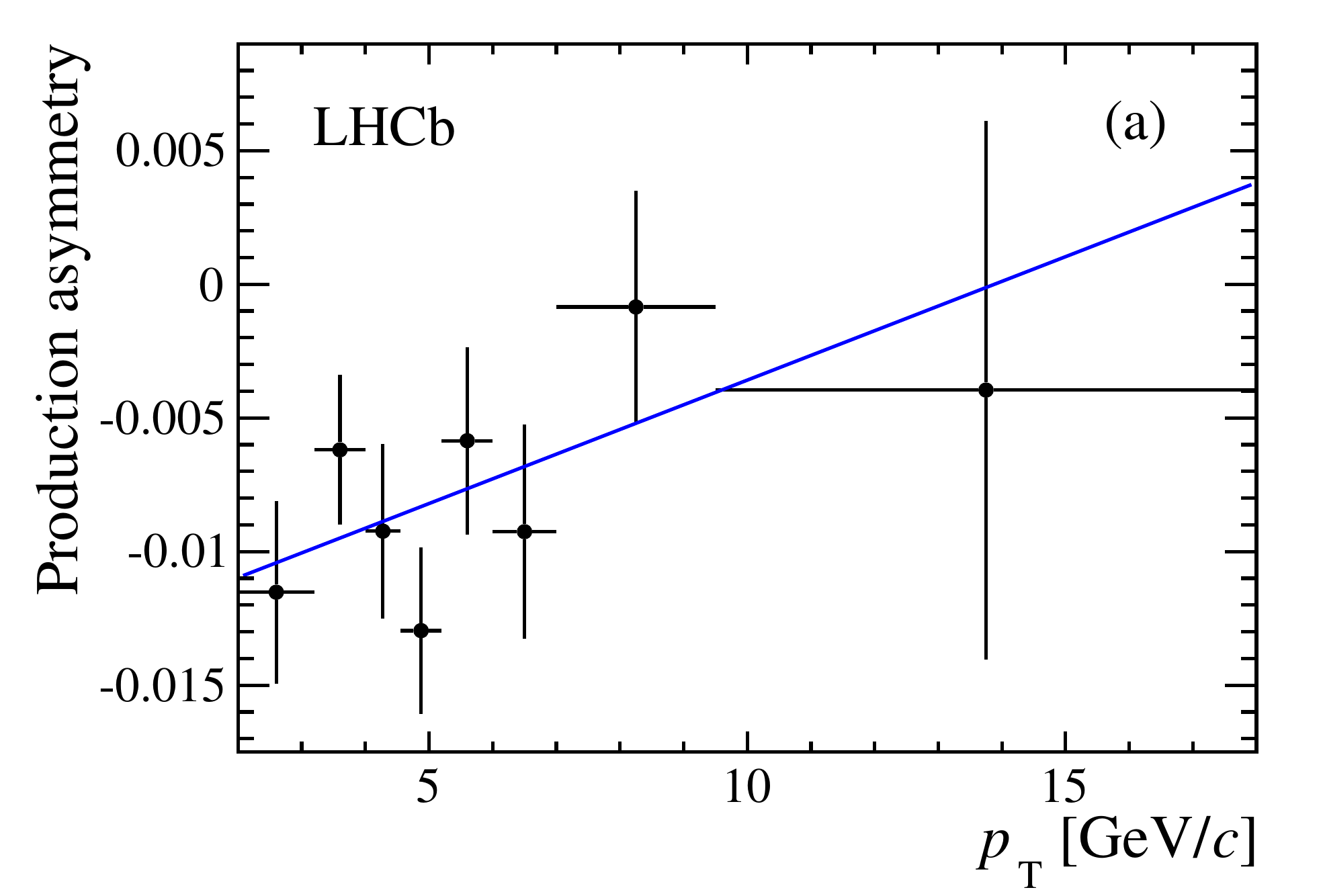}
\includegraphics[width=7cm]{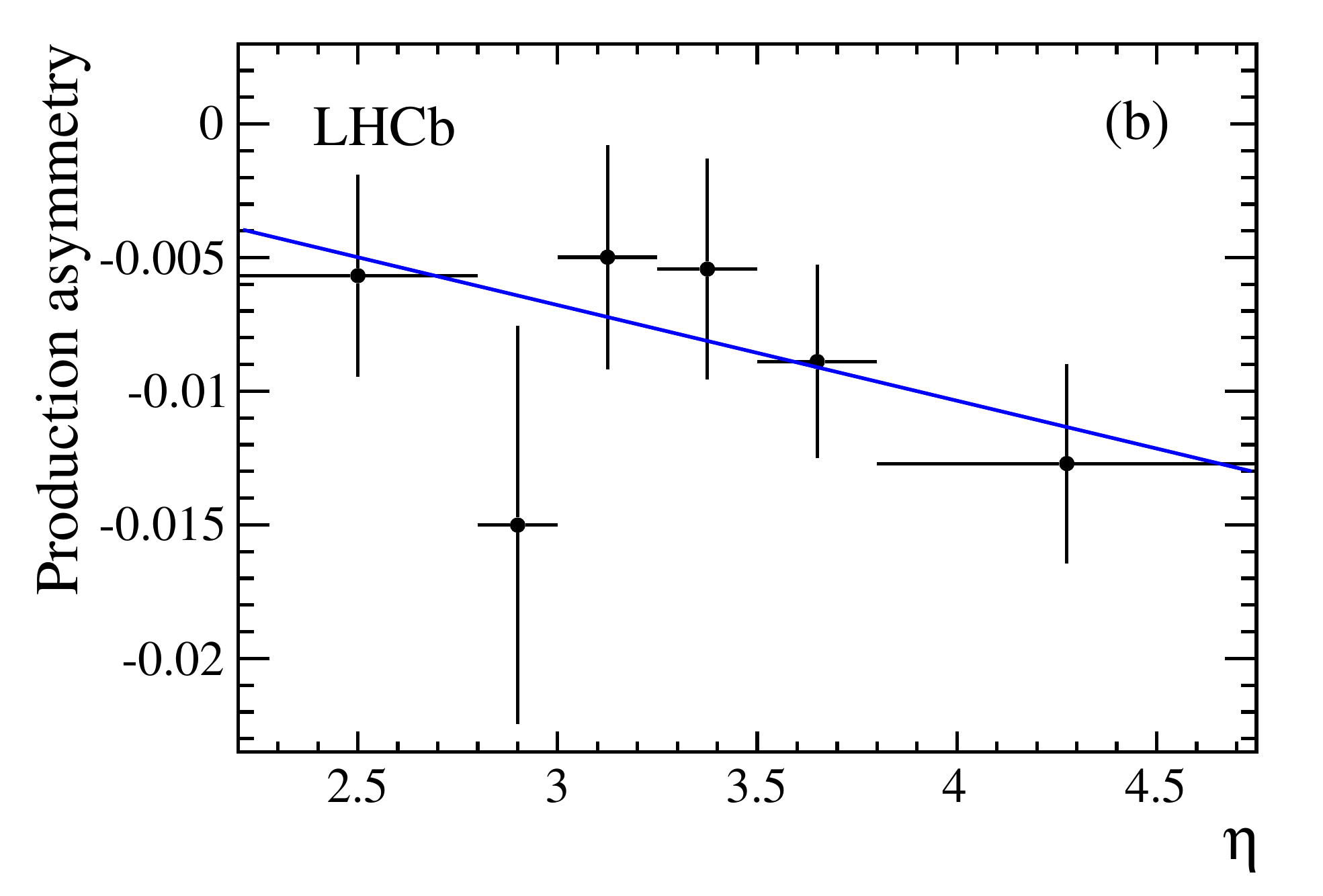}
\end{center}
\caption{Dependence of the $D^+$ production asymmetry on $p_{\rm T}$
  (left) and $\eta$ (right).\label{fig:dpluspteta}}
\end{figure}

 The average
  asymmetries are
\begin{equation}
A_{prod}(D_{(s)}^+) = (-0.33\pm0.13\pm0.18\pm0.10)\%
\end{equation}
\begin{equation}
A_{prod}(D^+) = (-0.96\pm0.19\pm0.18\pm0.18)\%
\end{equation}
where the uncertainties are statistical on the $D_{(s)}^{+}$ decays,
statistical on the pion efficiency asymmetry correction, and
systematic. There are some hints of the expected dependence on $p_{\rm
  T}$ and $\eta$ in the $D^+$ case, as shown in
Fig.~\ref{fig:dpluspteta}.

\begin{figure}
\begin{center}
\includegraphics[width=9cm]{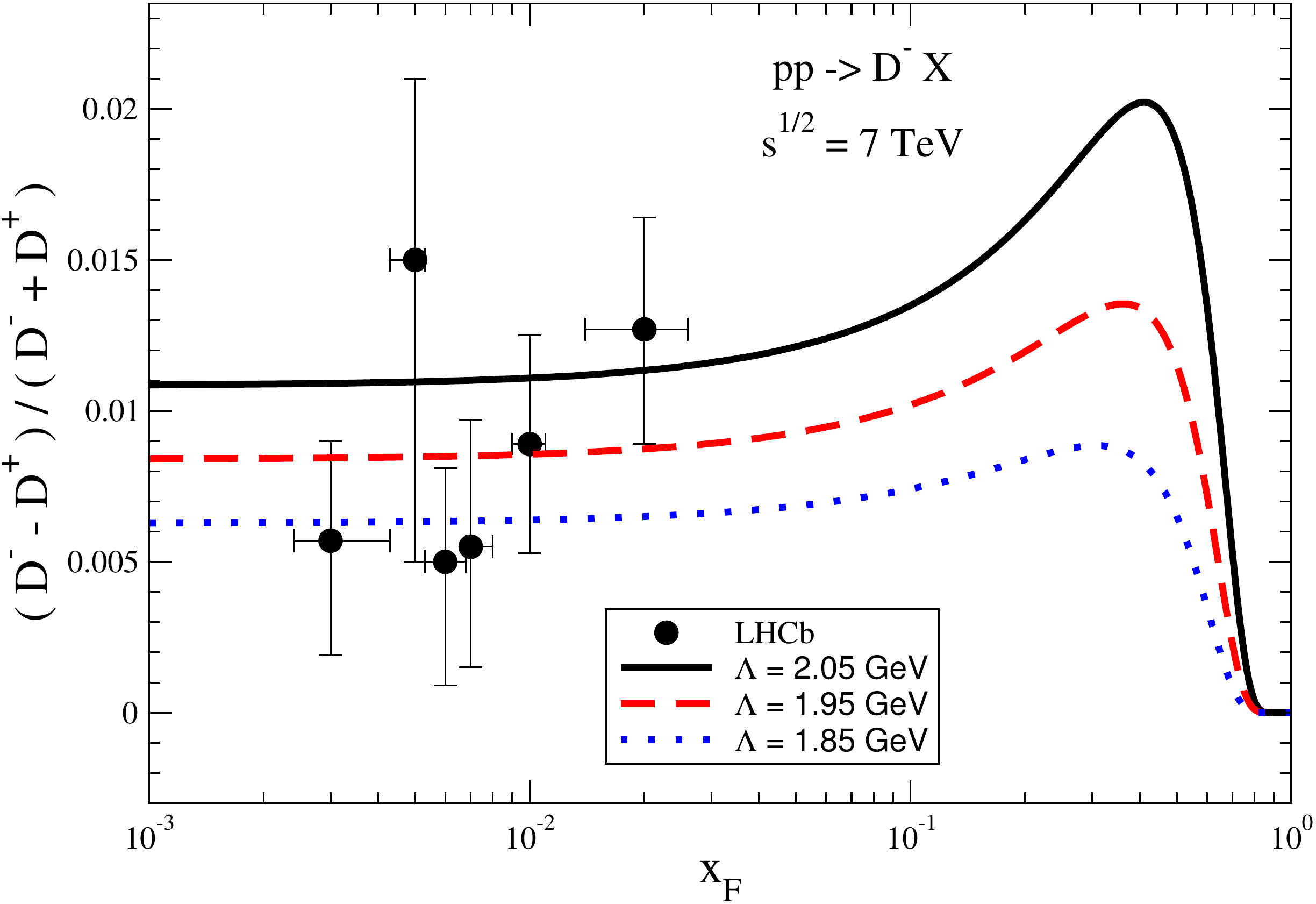}
\end{center}
\caption{Comparison of production asymmetries measured at LHCb with the
  predictions of the meson cloud theory~\cite{mesoncloud}. The parameter $\Lambda$ is a cut-off. Note that the
  opposite convention is used here to define asymmetry, with an excess of
  $D^-$ decays being defined as positive.\label{fig:theory}}
\end{figure}

The comparison of the results with the theoretical model of
Ref.~\cite{mesoncloud} is shown in Fig.~\ref{fig:theory}. It is clear
that the effect is relatively small and the dependence on kinematic
variables relatively weak, so more precise data will be needed before a
fully rigorous test of the theory can be performed.

\section{Experimental challenges}
\label{sec:sys}

As data samples get larger and larger, systematic uncertainties are
becoming increasingly important.  It is a generally held view that systematics can be controlled
  at the level of the statistical uncertainty, but to achieve this
  they must be studied in ever more detail.

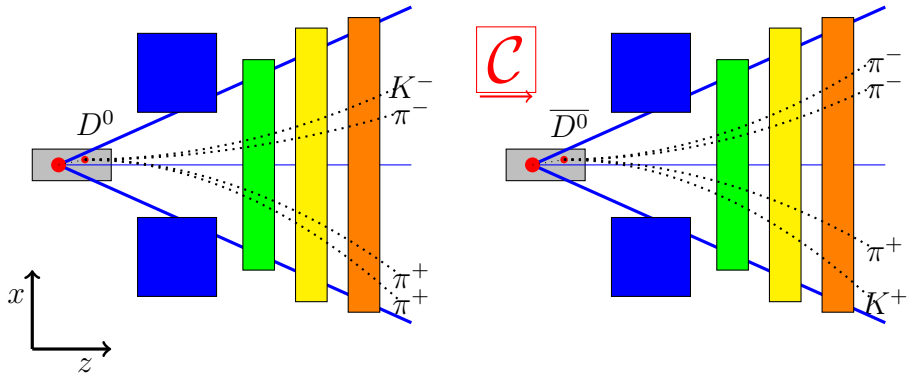
\begin{figure}
\begin{center}
\begin{tikzpicture}[scale=0.7,
			level/.style={level distance=3.15cm, line width=0.4mm},
			level 2/.style={sibling angle=60},
			level 3/.style={sibling angle=60},
			level 4/.style={level distance=1.4cm, sibling angle=60}
	]
        \node[draw=none,fill=none] at (-7.80, 0.8){$D^0$} ;
	\node[draw=none,fill=none] at (-1.8, 1.5){$K^-$} ;
        \node[draw=none,fill=none] at (-1.8, 1.0){$\pi^-$} ;
        \node[draw=none,fill=none] at (-1.8, -2.6){$\pi^+$} ;
        \node[draw=none,fill=none] at (-1.8, -2.1){$\pi^+$} ;
       \filldraw[draw=black,fill=lightgray] (-9.0,-0.3) rectangle
  (-7.5,0.3);
	\draw[blue,very thick] (-8.5,0.0) -- (-1.8,3.0) ;
        \draw[blue,very thick] (-8.5,0.0) -- (-1.8,-3.0) ;
        \draw[blue] (-8.5,0.0) -- (-1.8,0.0) ;
\draw[very thick, ->] (-9, -3.5) -- (-7.5, -3.5);
\draw[very thick, ->] (-9, -3.5) -- (-9, -2.0);
\node[draw=none,fill=none] at (-9.30, -2.5){$x$} ;
\node[draw=none,fill=none] at (-8.0, -3.8){$z$} ;

        \node[circle, fill=red,inner sep=1pt,minimum size=1pt] at
        (-8.0, 0.1){};
        \node[circle, fill=red,inner sep=2pt,minimum size=2pt] at (-8.5, 0.0){};
        
        \filldraw[draw=black,fill=blue] (-7.0,2.5) rectangle (-5.5,1.0);
        \filldraw[draw=black,fill=blue] (-7.0,-2.5) rectangle
        (-5.5,-1.0);
        \filldraw[draw=black,fill=green] (-5.0,-2.0) rectangle
        (-4.4,2.0);
        \filldraw[draw=black,fill=yellow] (-4.0,-2.6) rectangle
        (-3.4,2.6);
        \filldraw[draw=black,fill=orange] (-3.0,-2.8) rectangle
        (-2.4,2.8);
 
\draw[dotted] (-8.5,0.0) -- (-8.0,0.1) ;
        \draw[dotted, thick] (-8.0,0.1) parabola(-2,-2.6);
        \draw[dotted, thick] (-8.0,0.1) parabola(-2,-2.1);
        \draw[dotted, thick] (-8.0,0.1) parabola(-2,1.0);
        \draw[dotted, thick] (-8.0,0.1) parabola(-2,1.5);

\node[draw=red,fill=none, font=\huge,red] at (0.0, 2.0){${\cal C}$} ;
\draw[thick,red, style=->] (-0.5,1.3) -- (0.5,1.3) ;
        \node[draw=none,fill=none] at (1.20, 0.8){$\bar{D^0}$} ;
	\node[draw=none,fill=none] at (7.2, 1.5){$\pi^-$} ;
	\node[draw=none,fill=none] at (7.2, 2.0){$\pi^-$} ;
        \node[draw=none,fill=none] at (7.2, -1.6){$\pi^+$} ;
        \node[draw=none,fill=none] at (7.2, -2.6){$K^+$} ;
       \filldraw[draw=black,fill=lightgray] (0.0,-0.3) rectangle
  (1.5,0.3);
	\draw[blue,very thick] (0.5,0.0) -- (7.2,3.0) ;
        \draw[blue,very thick] (0.5,0.0) -- (7.2,-3.0) ;
        \draw[blue] (0.5,0.0) -- (7.2,0.0) ;

        \node[circle, fill=red,inner sep=1pt,minimum size=1pt] at
        (1.1, 0.1){};
        \node[circle, fill=red,inner sep=2pt,minimum size=2pt] at (0.5, 0.0){};
        
        \filldraw[draw=black,fill=blue] (2.0,2.5) rectangle (3.5,1.0);
        \filldraw[draw=black,fill=blue] (2.0,-2.5) rectangle
        (3.5,-1.0);
\filldraw[draw=black,fill=green] (4.0,-2.0) rectangle
(4.6,2.0);
        \filldraw[draw=black,fill=yellow] (5.0,-2.6) rectangle
        (5.6,2.6);
        \filldraw[draw=black,fill=orange] (6.0,-2.8) rectangle
        (6.6,2.8);
 
\draw[dotted] (0.5,0.0) -- (1.1,0.1) ;
        \draw[dotted, thick] (1.1,0.1) parabola(7,-2.6);
        \draw[dotted, thick] (1.1,0.1) parabola(7,-1.6);
        \draw[dotted, thick] (1.1,0.1) parabola(7,1.5);
        \draw[dotted, thick] (1.1,0.1) parabola(7,2.0);
\end{tikzpicture}
\end{center}
\caption{Schematic of the LHCb detector showing the path of charged
  particles from a $D^0 \to K^-\pi^+\pi^-\pi^+$ decay and its charge
  conjugate. In this case, the raw asymmetry will be dominated by the
  material interaction effects of the charged kaon, but when the pion
  tracking efficiency is measured, this cancels
  between the numerator and the denominator.\label{fig:detcartoon}}
\end{figure}

In charge asymmetry measurements,
important systematic uncertainties arise from
the fact that the magnetic field used to separate the charges bends oppositely-charged
particles in opposite directions so they pass through different parts
of the detector. The different detector elements could have
different efficiencies. This is illustrated for the LHCb detector in
Fig.~\ref{fig:detcartoon}. The different acceptance and efficiency in
different radial directions is responsible for the large
asymmetries seen in, for example, the pion detection efficiency as a
function of azimuthal angle in
Fig.~\ref{fig:pioneffic} for data taken with one magnet polarity. This figure highlights the importance of taking data
with both magnet polarities and averaging the results, as this leads
to near-complete cancellation of the effects.

 Other key systematic uncertainties in LHCb production and CP asymmetry
  measurements are associated with material interaction effects. The asymmetric
  interaction of positive and negative pions with detector material are responsible for most of the angle-independent asymmetry in
Fig.~\ref{fig:pioneffic}. Charged kaon material interactions lead to still
larger asymmetries. There are also nuisance effects from neutral kaon mixing and CP
  violation. Neutral kaons are particularly interesting
  because they violate CP and their mixing can be affected by material interactions. 

To parameterise neutral kaon material interactions, one usually defines a `regeneration parameter' $r$ in terms of forward scattering
  amplitudes $f$ and $\bar{f}$ for $K^0$ and $\bar{K^0}$ respectively,
\begin{equation}
r = -\frac{\pi {\cal N}(f - \overline{f})}{\Delta m - \frac{i}{2}(\Gamma_L-\Gamma_S)}
\end{equation}
where $ {\cal N}$ is the number density of atoms in the material,
$\Delta m $ is the mass difference between $K^0$ and $\bar{K^0}$, and
$\Gamma_{L,S}$ are their lifetimes. The imaginary part of $f$ is
related to the cross section by the optical theorem and the real part
of $f$ is related to the imaginary part by dispersion
integrals~\cite{dispint}. The difference between $K^0$ and $\bar{K^0}$
scattering amplitudes follows the scaling law
\begin{equation}
f-\bar{f} \propto -\frac{23.2pA^{0.758}}{[p\; ({\rm GeV}/c)]^{0.614}}\;{\rm mb}\end{equation} 
where $A$ is the nucleon number of the material and $p$ is the
momentum of the neutral kaon~\cite{scaling}.  Ko \emph{et al}~\cite{KoWon} model a detector as a series of layers of
    material, calculate $r$ for each layer using measured cross
    sections, and solve a set of
    recursive equations to determine the asymmetry as a function of
    the kaon decay time and momentum. The CPV in the neutral kaon
    system decouples from this regeneration at first order. Neglecting
    direct CPV, it is given by
\begin{equation}
A(t) = 2{\rm Re} (\epsilon) - 2e^{-\frac{1}{2}\Delta\Gamma t}\left({\rm Re} (\epsilon) \cos \Delta m t +
   {\rm Im} (\epsilon) \sin \Delta m t\right)
\end{equation}
where the indirect CP violation parameter $\epsilon$ is approximately $2\times10^{-3}$.

\begin{figure}
\begin{center}
\includegraphics[width=9cm]{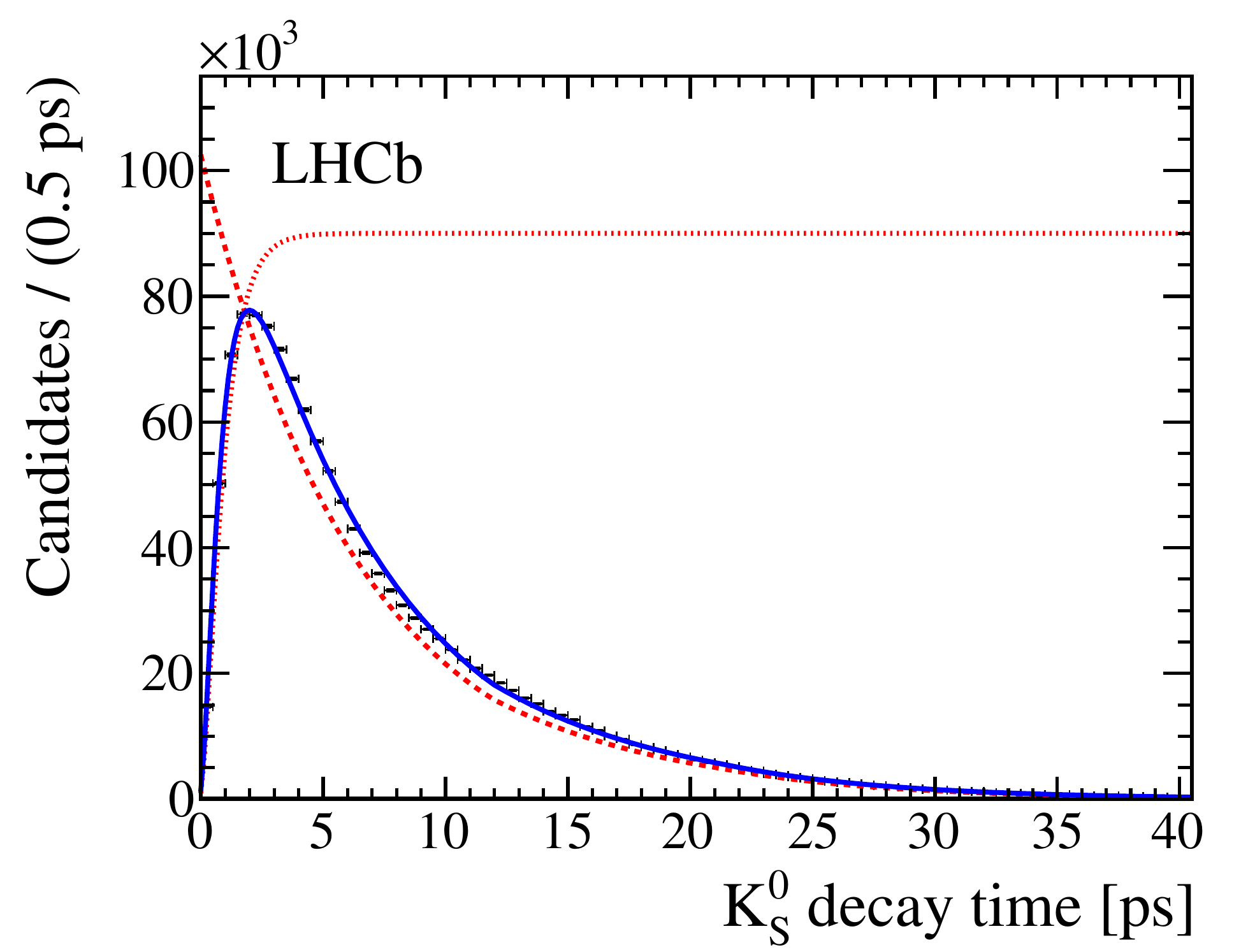}
\end{center}
\caption{The $K^0_S$ decay time distribution for neutral kaons
  selected for use in the production asymmetry
  analysis.~\label{fig:kslt}}
\end{figure}

 The formalism has now been  employed at LHCb, but to date 
 both material interactions and CPV lead to small effects on the
 measured raw asymmetries in charm decays of a few times
$10^{-4}$. This is because kaons used in  current analyses are very short-lived 
compared to $K^0_S$ lifetime of 89~ps, due to peculiarities in the
trigger and selection criteria. The decay time distribution of the kaons is shown
in Fig.~\ref{fig:kslt}.

\section{Perspective}

   With production asymmetries under control, it is possible to search
    for CP-violation more precisely and in more different ways.
    Sometimes one can extract the production and CPV asymmetries
  together, as done in the analysis of $D^+ \to K^0_S h^+$ by the
  Belle collaboration. Production asymmetries are also interesting for QCD and those
  measured in $pp$ collisions should
  help theorists to develop non-perturbative models of the proton. 

The prospects for the future include measurements of the
  $\Lambda_{c}^{+}$ and $D^{*+}$ production asymmetries at LHCb. These
  are likely to be challenging but rewarding analyses and the results
  will be highly pertinent to our understanding of both particle production
  and CP violation in charm decays.

\end{document}